\documentclass[a4paper,10pt]{article}
\pdfoutput=1 

\usepackage{jcappub} 

\usepackage[T1]{fontenc} 
\usepackage{orcidlink}

\title{\boldmath Geodesics motion of test particles around Schwarzschild-Klinkhamer wormhole with topological defects and gravitational lensing}

\author{Faizuddin Ahmed\,\orcidlink{0000-0003-2196-9622}}

\affiliation{Department of Physics, University of Science \& Technology Meghalaya, Ri-Bhoi, 793101, India}

\emailAdd{faizuddinahmed15@gmail.com; faizuddin@ustm.ac.in}

\abstract{This study investigates the geodesic motion of test particles, both massless and massive, within a Schwarzschild-Klinkhamer (SK) wormhole space-time. We specifically consider the influence of cosmic strings on the system and analyze the effective potential, and observing that the presence of a cosmic string parameter alters it for null and time-like geodesics. Moreover, we calculate the deflection angle for null geodesics, and demonstrate that the cosmic string modifies this angle and induces a shift in the results. Additionally, we extend our investigation in this SK-wormhole space-time but with a global monopole. We explore the geodesic motion of test particles in this scenario and find that the effective potential is affected by the global monopole. Similarly, we determine the deflection angle for null geodesics and show that the global monopole parameter introduces modifications to this angle. Lastly, we present several known solutions for space-times involving cosmic strings and global monopoles within the framework of this SK-wormhole.}


\arxivnumber{2307.08503}

\begin{document}
\maketitle
\flushbottom

\section{Introduction}
\label{sec:intro}

Geodesic motion plays a crucial role in understanding the fundamental aspects of black hole (BH) and wormhole space-times. It provides valuable insights into the underlying structure of the space-time and offers significant information pertaining to these astrophysical phenomena. The behavior of geodesics in the background space-time exhibits a remarkable complexity, carrying essential implications for BHs or wormholes. Among the various types of geodesic motions, circular geodesic motion holds particular interest due to its connection with gravitational binding energy. These motions provide insights into the dynamics of particles within the gravitational field. Notably, null geodesics are highly effective in explaining characteristic modes of BHs, such as quasinormal modes, as well as the behavior of wormholes. In the case of an extreme Schwarzschild black hole, there exist a simple set of unstable circular orbits, which arise as a consequence of the non-linear nature of GR. These orbits represent an important outcome of the theory, shedding light on the intricate dynamics of gravitational systems.

Gravitational lensing \cite{ab1,ab2} has been a prominent area of research in cosmology and gravitation ever since its initial experimental observation by A. Eddington following the proposal of the general theory of relativity in 1915. This phenomenon manifests in both weak field scenarios, where light rays pass at a considerable distance from the source, and strong field situations, where light rays come into close proximity with massive objects. Extensive investigations have been undertaken to explore gravitational lensing, encompassing a wide array of space-time configurations. These studies encompass diverse subjects such as gravitational lensing by black holes \cite{aa2,aa4,aa10,aa13,aa35,aa37,aa39,bb1,cc1,cc3,cc4,cc5,cc6,gg1}, naked singularities \cite{cc7,aa6,aaa6,aa8,aaa8}, spherically symmetric space-time \cite{aa15}, rotating global monopole space-times \cite{aa14,aa24}, cosmic string space-time under LSV \cite{aa19}, including the cosmological constant in rotating cosmic string space-time \cite{aa22}, Kerr-Newman-Kasuya space-time \cite{aa23}, black-bounce space-time \cite{JRN,cc2}, stationary axisymmetric space-time \cite{aa38}, and more. Furthermore, gravitational lensing has been investigated in topologically charged space-times within the framework of Eddington-inspired Born-Infeld gravity \cite{ARS}, massive gravity \cite{aa44}, spherically symmetric and static space-time \cite{aa26}, topologically charged Eddington-inspired Born-Infeld gravity space-time \cite{aa33}, and a space-time with cosmic string in Eddington-inspired Born-Infeld gravity space-time \cite{hh1}.

Wormholes are hypothetical passageways that connect different regions of the same universe or even distinct universes through a narrow throat. The study of wormholes originated from the manipulation of the vacuum solution of the Schwarzschild black hole. However, in 1973, the exact solution for a wormhole was first provided by Ellis \cite{HGE}, with a similar solution independently proposed by Bronnikov \cite{KAB} in the same year. These wormholes, collectively known as the Ellis-Bronnikov wormhole, have been extensively explored in scientific literature. The fascination with wormholes grew significantly following the groundbreaking works of Morris and Thorne on a similar type of wormhole referred to as the EB-type wormhole \cite{MT}. This particular wormhole concept introduced a novel perspective on the possibility of time travel within the framework of general relativity theory. The origins of this concept can be traced back to 1949 when K. Gödel presented his seminal work on a rotating universe \cite{GO}. Numerous wormhole solutions, both with and without a cosmological constant, have been constructed and investigated in scientific literature, contributing to our understanding of these hypothetical structures (for a comprehensive review, see Ref. \cite{MV}). Gravitational lensing in various wormholes geometries has been extensively studied by numerous authors \cite{RS, RS2, aa36, aa16, aa21, aa25, aa27, aa18, aa28, aa20, aa40, aa411, aa43, aa42, FB, FB2, aa29, aa31, aa30, aa41, bb4, bb5, bb6}. These investigations have shed light on the gravitational lensing effects and properties associated with wormholes.

Topological defects are theoretical entities that were believed to have formed during the early stages of the universe's evolution, as discussed comprehensively in the review articles \cite{kk1,kk2,kk3,AV,kk5,kk6}. These defects could have emerged during one of the early phase transitions in particle physics models, such as those associated with the breaking of grand unification symmetry or spontaneously broken gauge theories, as proposed in numerous field theory models \cite{kk7,kk8,kk9}. If the scale of symmetry breaking $\eta$ in the theory is on the order of $10^{16}$ GeV and the mass per unit length $\mu$ of the cosmic strings satisfies $G\,\mu/c^2 \approx 10^{-6}$, then these strings offer a mechanism for explaining the large-scale structures observed in the cosmos. The possibility of directly observing these cosmic strings through the gravitational lensing effect was suggested in \cite{kk0}, and there has been ongoing research into detecting their existence (for a review of the experimental status, see \cite{kk10,kk11,kk110}). Cosmic strings, if they do indeed exist, span immense cosmological distances and, despite being extremely thin, possess enough mass to exert noticeable gravitational effects.

It is worth noting that strings can also manifest in theories devoid of gauge fields, where the symmetry is global rather than local gauge symmetry. These are referred to as global monopoles, and they take the form of spherically symmetric objects arising from the self-coupling triplet of scalar fields $\phi^{a}$. These scalar fields undergo a spontaneous breaking of the global $O(3)$ gauge symmetry, resulting in structures akin to cosmic strings with $U(1)$ symmetry. However, global monopoles exhibit distinct properties. Notably, studies have demonstrated that these topological defects possess a negative gravitational potential \cite{kk12}. The concept of global monopole charge has also found relevance in cosmological contexts, as evidenced by research in \cite{kk13,kk14,kk15}. Moreover, recent investigations have delved into the existence of wormholes within the Milky Way galaxy, taking into account the presence of global monopole charge, as discussed in \cite{kk16}.

In Ref. \cite{ss1}, the author presented a new type of wormhole geometry, a non-vacuum solution of the Einstein's field equations which is a degenerate one. This new wormhole is nomenclature as defect wormhole or Klinkhamer-defect wormhole without exotic matter. The line-element describing this defect wormhole in the chat $(t, x, \theta, \phi)$ is given by
\begin{equation}
    ds^2=-dt^2+\Big(\frac{x^2}{x^2+b^2}\Big)\,dx^2+(x^2+a^2)\,(d\theta^2+\sin^2 \theta\,d\phi^2).
    \label{K1}
\end{equation}
Here $a, b$ are nonzero real constants and the coordinates $-\infty < (t, x) <+\infty$, $\theta \in [0,\pi)$, and $\phi \in [0, 2\,\pi]$. There author have shown that matter-distribution satisfies the weak energy condition (WEC) and the null energy condition (NEC) provided the parameters is related by $b^2 > a^2$. The geometrical and curvature properties of this defect wormhole have been discussed in details there. In Ref. \cite{sss1}, the same author extensively explores a vacuum-defect wormhole space-time derived by setting $b^2=a^2$ within the framework of the previously mentioned defect wormhole metric (\ref{K1}). Notably, it's important to highlight that in a separate study in \cite{JCF}, an analysis is presented indicating that smooth metrics can effectively conceal the presence of thin shells. Furthermore, an additional commentary on defect wormholes is provided in \cite{BGV}. The defect wormhole solution remains a pertinent topic within the realm of nonstandard general relativity. In this context, the presence of a defect at the point $x=0$ signifies an inherent imperfection within the fabric of space-time. It is at this juncture that the conventional elementary-flatness condition, which characterizes smooth and unblemished space-time, fails to hold true. Regrettably, it seems that this crucial aspect has been inadvertently overlooked in the more recent papers, namely, the works of \cite{JCF} and \cite{BGV}.

In Ref. \cite{ss3}, author discussed the thermodynamics of Schwarzschild black hole with a cosmic string. In the coordinate chat $(t, r, \theta, \phi)$ with $0 \leq r$ and $0 \leq \phi < 2\,\pi$, the line-element is described by 
\begin{equation}
    ds^2=-A(r)\,dt^2+\frac{dr^2}{A(r)}+r^2\,(d\theta^2+\alpha^2\,\sin^2 \theta\,d\phi^2),\quad A(r)=\Big(1-\frac{2\,M}{r}\Big).
    \label{K2}
\end{equation}
One can see that in the limit $M \to 0$, this line-element (\ref{K2}) reduces to a cosmic string space-time in the spherical system given by $ds^2=-dt^2+dr^2+r^2\,(d\theta^2+\alpha^2\,\sin^2 \theta\,d\phi^2)$ \cite{AV,GAM}.

In this study, our objective is to examine the behavior of light-like and time-like geodesics within the framework of a Schwarzschild-Klinkhamer (SK) defect wormhole, taking into account the presence of topological defects caused by a cosmic string and global monopole. Specifically, we investigate the effective potential of the system in both geometries and analyze how the parameters of cosmic string and global monopole influences this potential, leading to a shift in the results obtained. Furthermore, we focus on determining the deflection angle for light-like geodesics within these geometries. We observe that the topological defects arising from the cosmic string and global monopole introduce modifications to this deflection angle, indicating that the presence of these defects has an impact on the bending of light within these wormhole. Additionally, we discuss various well-known space-time geometries that can be obtained as special cases from this SK-wormhole with topological defects, further highlighting the significance of these topological defects in shaping the overall properties of these wormholes.

\section{Schwarzschild-Klinkhamer wormhole with a cosmic string}

In this section, we consider a generalized version of the Klinkhamer wormhole space-time (\ref{K1}) by incorporating a factor $A(x)$ analogue to the Schwarzschild solution in the presence of a cosmic string passing through it. This new metric we called Schwarzschild-Klinkhamer wormhole space-time with a cosmic string, a non-vacuum solution of the field equations. This new metric is described by the following line-element in the chat $(t, x, \theta, \phi)$ as \footnote{After this work, we noticed a paper \cite{ff1}, where a similar metric without topological defects, such as cosmic string or global monopole charge was presented.}
\begin{eqnarray}
      ds^2=-A(x)\,dt^2+\frac{dx^2}{\Big(1+\frac{b^2}{x^2}\Big)\,A(x)}+(x^2+a^2)\,(d\theta^2+\alpha^2\,\sin^2 \theta\,d\phi^2),\,
      A(x)=\Big[1-\frac{2\,M}{\sqrt{x^2+b^2}}\Big].
\label{a1}
\end{eqnarray}
Here $0 < \alpha \leq 1$ is a cosmic string parameter related with linear mass density of the string. We choose the parameters $a, b$, and $M$ all are positive with $b > 2\,M$, $b^2 > a^2$, and $a>0$. The above Schwarzschild-Klinkhamer wormhole metric (\ref{a1}) can reduce to Schwarzschild solution with different spatial structure, and the original Klinkhamer wormhole metric (\ref{K1}) \cite{ss1}. Note that the condition $b>2\,M$ in (\ref{a1}) is required to prevent the formation of event horizon. The introduction of a cosmic string into the space-time brings about significant alterations to its geometrical and curvature characteristics.

The nonzero components of the Einstein tensor $G_{\mu\nu}$ for the space-time (\ref{a1}) are 
\begin{eqnarray}
    -G^{t}_{t}=\frac{(b^2-a^2)}{(x^2+a^2)^2}\,\Big(1-\frac{2\,M}{\sqrt{x^2+b^2}}\Big)=G^{x}_{x},\quad G^{\theta}_{\theta}=G^{\phi}_{\phi}=-\frac{R}{2}.
    \label{tensor}
\end{eqnarray}
The Ricci scalar $R$ and the Kretschmann scalar curvatures of the above space-time are given by
\begin{eqnarray}
    &&R=\frac{4\,M}{(x^2+b^2)^{3/2}}-\frac{2\,(a^2-b^2+2\,M\,\sqrt{x^2+b^2})}{(x^2+a^2)^2},\nonumber\\
    &&\mathcal{K}=\frac{4}{(x^2+a^2)^4\,(x^2+b^2)^3}\,\Big[4\,M^2\,(a^2+x^2)^4-4\,M\,(a^2 + x^2)^2\,(b^2 + x^2)^{5/2}- 
 12\,b^2\,M\,(b^2 + x^2)^{7/2}\nonumber\\
 &&-12\,M\,x^2 (b^2 + x^2)^{7/2}+16\,M\,(a^2 + x^2)\,(b^2 + x^2)^{7/2}+3\,(b^2 + x^2)^4\,(b^2 + 4 M^2 + x^2)\nonumber\\
 &&+ (a^2 + x^2)^2\,(b^2 + x^2)^2\,(3 b^2 + 4 M^2 + 3 x^2)-2\,(a^2 + x^2)\,(b^2 + x^2)^3\,(3 b^2 + 4 M^2 + 3 x^2) \Big].
    \label{tensor2}
\end{eqnarray}

For the energy-momentum tensor $T^{\mu}_{\nu}=\mbox{diag}\,(-\rho, p_{x}, p_{t}, p_{t})$, where $\rho$ is the energy-density, and $p_x, p_t$ are the pressure components along $x$- and tangential directions, we obtain  
\begin{eqnarray}
    &&\rho=p_x=\frac{(b^2-a^2)}{(x^2+a^2)^2}\,\Big(1-\frac{2\,M}{\sqrt{x^2+b^2}}\Big)>0,\nonumber\\
    &&p_t=-\frac{2\,M}{(x^2+b^2)^{3/2}}+\frac{(a^2-b^2+2\,M\,\sqrt{x^2+b^2})}{(x^2+a^2)^2}    
\end{eqnarray}
provided $b^2 > a^2$, where $b>2\,M$ to satisfies the weak energy condition. We see that these physical quantities and the scalar curvatures defined in (\ref{tensor2}) are finite at $x=0$ and vanishes for $x \to \pm\,\infty$.

To check the validate of the energy conditions, we consider the following vector fields, such as the null vector $k^{\mu}$, the time-like four-vector $U^{\mu}$, and a spacelike unit vector $\eta^{\mu}$ along $x$-direction for the space-time (\ref{a1}) as
\begin{eqnarray}
    &&k^{\mu}=\frac{1}{\sqrt{2}}\,\Bigg[-\Big(1-\frac{2\,M}{\sqrt{x^2+b^2}}\Big)^{-1/2}, \sqrt{1+\frac{b^2}{x^2}}\,\Big(1-\frac{2\,M}{\sqrt{x^2+b^2}}\Big)^{1/2}, 0, 0\Bigg],\nonumber\\
    &&U^{\mu}=\Big(1-\frac{2\,M}{\sqrt{x^2+b^2}}\Big)^{-1/2}\,\delta^{\mu}_{0},\nonumber\\
    &&\eta^{\mu}=\Bigg[\Big(1-\frac{2\,M}{\sqrt{x^2+b^2}}\Big)^{1/2}\,\sqrt{1+\frac{b^2}{x^2}}\Bigg]\,\delta^{\mu}_{x},
    \label{mnull}
\end{eqnarray}
where these vector fields satisfy the following relations
\begin{equation}
    -U^{\mu}\,U_{\mu}=1=\eta^{\mu}\,\eta_{\mu},\quad U_{\mu}\,\eta^{\mu}=0=k^{\mu}\,k_{\mu},\quad k_{\mu}\,\eta^{\mu}=\frac{1}{\sqrt{2}}=U_{\mu}\,k^{\mu}.\label{mnull1}
\end{equation}

We choose the energy-momentum tensor of the following form 
\begin{eqnarray}
    T_{\mu\nu}=(\rho+p_{t})\,U_{\mu}\,U_{\nu}+p_{t}\,g_{\mu\nu}+(p_{x}-p_{t})\,\eta_{\mu}\,\eta_{\nu}.
    \label{mnull2}
\end{eqnarray}
With the help of (\ref{mnull}), we find the following energy conditions
\begin{equation}
    T_{\mu\nu}\,U^{\mu}\,U^{\nu}=\rho>0\quad,\quad T_{\mu\nu}\,k^{\mu}\,k^{\nu}=\frac{1}{2}(\rho+p_x)>0
    \label{mnull3}
\end{equation}
provided $b^2 > a^2$. 

Thus, from the above analysis it is clear that the wormhole space-time considered by (\ref{a1}) satisfies the weak energy condition (WEC) and the null energy condition (NEC).

Now, we discuss below a few special cases of the above Schwarzschild-Klinkhamer wormhole metric with a cosmic string.

\vspace{0.2cm}

{\bf CASE A}: If one choose $M \to 0$, then the space-time (\ref{a1}) reduces to the following form
\begin{equation}
      ds^2=-dt^2+\frac{1}{\Big(1+\frac{b^2}{x^2}\Big)}\,dx^2+(x^2+a^2)\,(d\theta^2+\alpha^2\,\sin^2 \theta\,d\phi^2).
\label{a2}
\end{equation}
The above metric (\ref{a2}) is called the Klinkhamer defect wormhole (\ref{K1}) with a cosmic string.

For the case $b^2=a^2$, the metric (\ref{a2}) reduces to the following form
\begin{equation}
    ds^2=-dt^2+\Big(1+\frac{b^2}{x^2}\Big)^{-1}\,dx^2+(x^2+b^2)\,(d\theta^2+\alpha^2\,\sin^2 \theta\,d\phi^2)
    \label{a7}
\end{equation}
which is the well-known Klinkhamer vacuum-defect wormhole \cite{sss1} with a cosmic string (see Eq. (5) in \cite{sss1}). Transforming to a coordinate via $r=\sqrt{x^2+b^2}$ into the metric (\ref{a7}), we obtain
\begin{equation}
    ds^2=-dt^2+dr^2+r^2\,(d\theta^2+\alpha^2\,\sin^2 \theta\,d\phi^2)
    \label{a8}
\end{equation}
which looks similar to a cosmic string space-time in the spherical system \cite{AV,GAM} with $r \in [b, \infty)$ instead of $r \in[0, \infty)$. It is worthwhile mentioning that the gravitational lensing effects of vacuum strings was investigated in Ref. \cite{JRG}.

\vspace{0.2cm}

{\bf CASE B}: If one choose $b^2=a^2$, where $2\,M < b < 3\,M$, then the metric (\ref{a1}) reduces to the following form 
\begin{eqnarray}
    ds^2=-\Big(1-\frac{2M}{\sqrt{x^2+b^2}}\Big)dt^2+\frac{\Big(1+\frac{b^2}{x^2}\Big)^{-1}}{\Big(1-\frac{2M}{\sqrt{x^2+b^2}}\Big)}\,dx^2+(x^2+b^2)(d\theta^2+\alpha^2 \sin^2 \theta d\phi^2)
    \label{a3}
\end{eqnarray}
which is a vacuum-defect wormhole with cosmic string since the Einstein tensor $G^{\mu}_{\,\nu}$ from equation (\ref{tensor}) will vanish. 

The Kretschmann scalar curvature of the metric given by $\mathcal{K}=\frac{48\,M}{(x^2+b^2)^2}$ is finite at $x=0$ and vanishes for $x \to \pm\,\infty$. Transforming to a new coordinate via $r=\sqrt{x^2+b^2}$ into the space-time (\ref{a3}), we obtain
\begin{eqnarray}
    ds^2=-\Big(1-\frac{2\,M}{r}\Big)\,dt^2+\Big(1-\frac{2\,M}{r}\Big)^{-1}\,dr^2+r^2\,(d\theta^2+\alpha^2\,\sin^2 \theta\,d\phi^2),
    \label{a4}
\end{eqnarray}
which looks similar form to the Schwarzschild vacuum solution with a cosmic string \cite{ss3} (see Eq. (5) in \cite{ss3}), where $r \in [b>2\,M, \infty)$ instead of $r \in [0, \infty)$.

\vspace{0.2cm}

{\bf CASE C}: If one choose $M \to 0$ and $b \to 0$, then the space-time (\ref{a1}) reduces to the following form
\begin{equation}
    ds^2=-dt^2+dx^2+(x^2+a^2)\,(d\theta^2+\alpha^2\,\sin^2 \theta\,d\phi^2),
\label{a5}
\end{equation}
an example of Morris-Thorne-type wormhole metric with a cosmic string \cite{aa31}, a non-vacuum solution of the Einstein's field equations. The Ricci scalar given by $R=-\frac{2\,a^2}{(x^2+a^2)^2}$, and the Kretschmann scalar curvature by $\mathcal{K}=\frac{12\,a^4}{(x^2+a^2)^4}$ both are finite at $x=0$ and vanishes for $x \to \pm\,\infty$.

\vspace{0.2cm}

{\bf CASE D}: If one choose $M \to 0$ and $a \to 0$, then the space-time (\ref{a1}) reduces to the following form
\begin{equation}
      ds^2=-dt^2+\Big(1+\frac{b^2}{x^2}\Big)^{-1}\,dx^2+x^2\,(d\theta^2+\alpha^2\,\sin^2 \theta\,d\phi^2),
\label{a6}
\end{equation}
another example of Morris-Thorne-type wormhole space-time with a cosmic string in Eddington-inspired Born-Infeld (EiBI) gravity background. This metric is a non-vacuum solution of the field equations with the Ricci scalar $R=\frac{2\,b^2}{x^4}$ and the Kretschmann scalar curvature $\mathcal{K}=\frac{12\,b^4}{x^8}$ both diverges at $x \to 0$ and vanishes for $x \to \pm\,\infty$. Recently, a space-time with cosmic string in Eddington-inspired Born-Infeld (EiBI) gravity background has been discussed in Ref. \cite{hh1}.

\section{Geodesics motion of test particles by wormhole with a cosmic string}

In this section, we analyze the motion of both massless and massive particles along geodesics in a wormhole background described by the metric (\ref{a1}). Notably, this space-time exhibits independence with respect to the coordinates $t$ and $\phi$, allowing for the existence of two Killing vectors. These Killing vectors are denoted as $\eta^{\mu}_{(t)}=\delta^{\mu}_{t}$, associated with time-translational invariance, and $\eta^{\mu}_{(\phi)}=\delta^{\mu}_{\phi}$, associated with rotational invariance. During the motion of a photon or time-like geodesics, the projection of the four-momentum of the particles along these Killing vectors remains conserved. This conservation gives rise to two fundamental quantities: the energy $E=-p_{\mu}\,\eta^{\mu}_{(t)}$, and the angular momentum $J=p_{\mu}\,\eta^{\mu}_{(\phi)}$.

We begin  this section by writing the Lagrangian of a system defined by \cite{aa6,aa8,aa25,aa28,aa30,aa31}
\begin{equation}
\mathcal{L}=\frac{1}{2}\,g_{\mu\nu}\,\left(\frac{dx^{\mu}}{d\tau}\right)\,\left(\frac{dx^{\nu}}{d\tau}\right),
\label{1}
\end{equation}
where $\tau$ is the affine parameter along the geodesics, and $g_{\mu\nu}$ is the metric tensor. Using the line-element (\ref{a1}), the Lagrangian (\ref{1}) in the equatorial plane $\theta=\frac{\pi}{2}$ (for photon or massive particles confined to the equatorial plane) is given by
\begin{equation}
\mathcal{L}=\frac{1}{2}\,\Bigg[-A(x)\,\dot{t}^2+\frac{1}{\Big(1+\frac{b^2}{x^2}\Big)}\,\frac{\dot{x}^2}{A(x)}+\alpha^2\,(x^2+a^2)\,\dot{\phi}^2\Bigg].
    \label{2}
\end{equation}
As stated earlier, there are two constant of motions and these are given by
\begin{eqnarray}
    A(x)\,\dot{t}=E\Rightarrow \dot{t}=\frac{E}{A(x)}\quad \mbox{and} \quad \alpha^2\,(x^2+a^2)\,\dot{\phi}=J\Rightarrow \dot{\phi}=\frac{J}{\alpha^2\,(x^2+a^2)}.
    \label{4}
\end{eqnarray} 

For null or time-like geodesics, the Lagrangian (\ref{2}) after transforming $r=\sqrt{x^2+b^2}$ becomes
\begin{equation}
    \frac{1}{2}\,\Big(\frac{dr}{d\tau}\Big)^2+V_{eff} (r)=\frac{E^2}{2}
    \label{5}
\end{equation}
which has the same form of the equation for a unit mass particle with energy $E^2/2$ moving in one-dimensional effective potential given by
\begin{equation}
    V_{eff} (r)=\Big(\frac{1}{2}-\frac{M}{r}\Big)\,\Bigg[\frac{J^2}{\alpha^2\,(r^2+a^2-b^2)}-\varepsilon\Bigg],
    \label{6}
\end{equation}
where $\varepsilon=0$ for null geodesics and $\varepsilon=-1$ for time-like geodesics.

\begin{figure}[tbp]
    \includegraphics[width=0.44\textwidth]{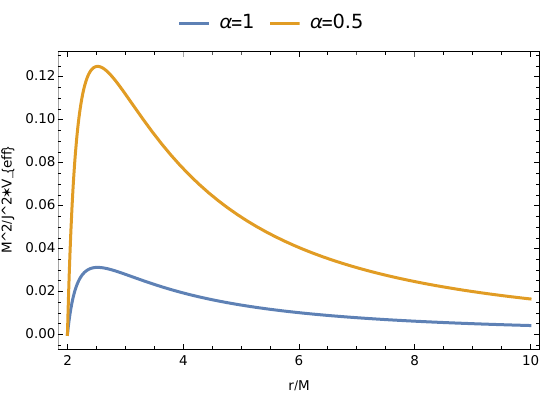}
    \hfill
    \includegraphics[width=0.44\textwidth]{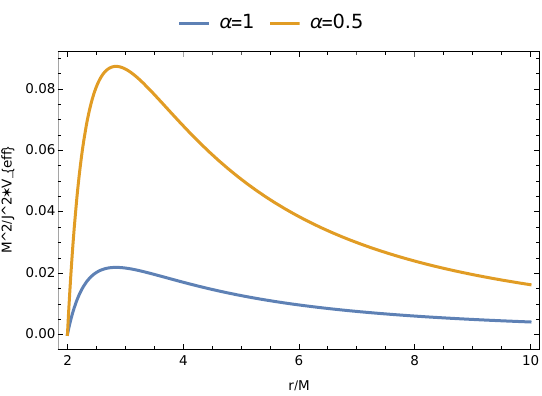}
\caption{\label{fig:1} Effective potential for null geodesics without ($\alpha=1$) and with cosmic string ($\alpha=0.5$) parameter. $a=1.5\,M$, $b=2.3\,M$ (left); $a=2\,M$, $b=2.3\,M$ (right) }
\end{figure}

\begin{figure}[tbp]
    \includegraphics[width=0.45\textwidth]{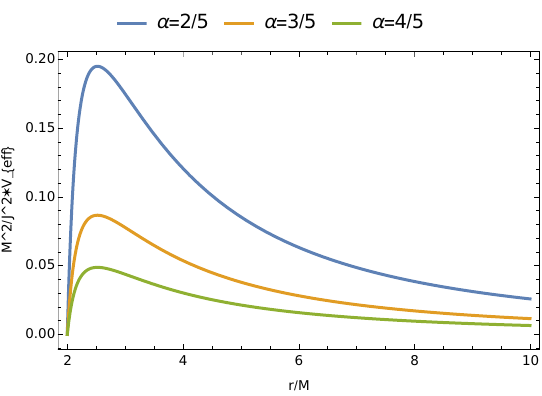}
    \hfill
    \includegraphics[width=0.45\textwidth]{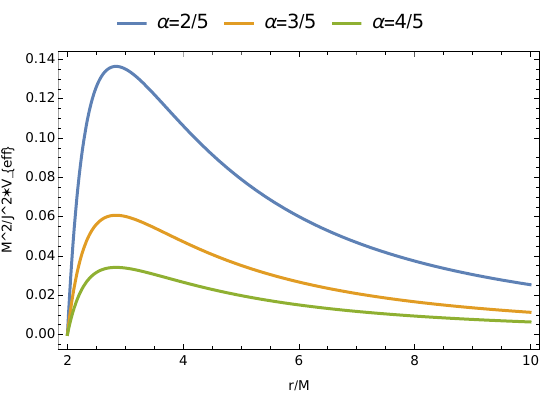}
    \caption{\label{fig: 2} Effective potential for null geodesics for different values of cosmic string parameter $\alpha$. $a=1.5\,M$, $b=2.3\,M$ (left); $a=2\,M$, $b=2.3\,M$ (right)}
\end{figure}

\begin{figure}[tbp]
    \includegraphics[width=0.45\textwidth]{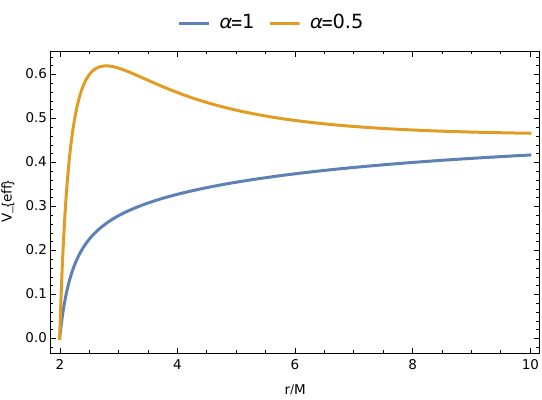}
    \hfill
    \includegraphics[width=0.45\textwidth]{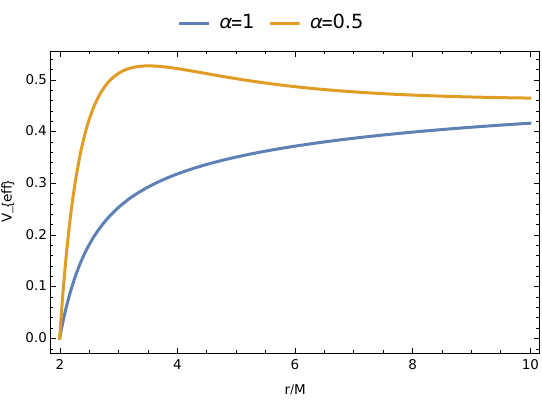}
    \caption{\label{fig: 3} Effective potential for time-like geodesics without ($\alpha=1$) and with cosmic string ($\alpha=0.5$) parameter. $a=1.5\,M$, $b=2.3\,M$, $J=2\,M$ (left); $a=2\,M$, $b=2.3\,M$, $J=2\,M$ (right) }
\end{figure}

\begin{figure}[tbp]
    \includegraphics[width=0.45\textwidth]{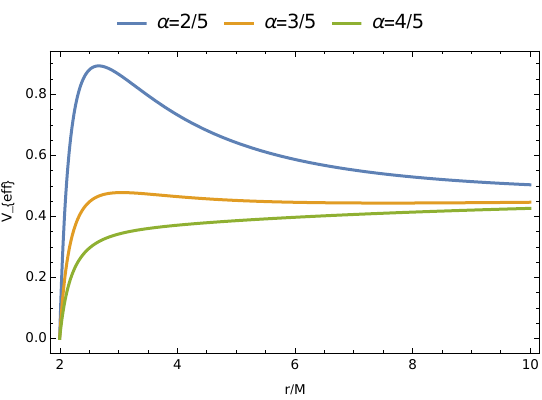}
    \hfill
    \includegraphics[width=0.45\textwidth]{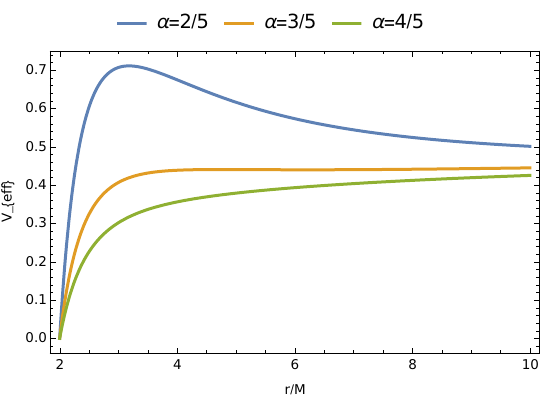}
    \caption{\label{fig: 4}Effective potential for time-like geodesics for different values of cosmic string parameter $\alpha$. Here, $a=1.5\,M$, $b=2.3\,M$, $J=2\,M$ (left); $a=2\,M$, $b=2.3\,M$, $J=2\,M$ (right). }
\end{figure}

The effective potential of the system described by equation (\ref{6}) is influenced not only by the parameters $a$ and $b$, but also by the cosmic string parameter $\alpha$. In gravitational and cosmological contexts, the cosmic string parameter typically lies within the interval $0 < \alpha < 1$. Consequently, the effective potential is increased due to the presence of cosmic string in the system.

In Figure 1, we present plots of the effective potential for null geodesics with different values of the cosmic string parameter: $\alpha = 1$ and $\alpha = 0.5$. On the left, we set $b = 2.3\,M>2\,M$ and $a = 1.5\,M$, while on the right, we use $b = 2.3\,M>2\,M$ and $a = 2\,M$. It is evident that the effective potential for null geodesics is significantly influenced by the cosmic string, showing a higher magnitude compared to the case without the cosmic string.

Moving on to Figure 2, we plot the effective potential for various values of the cosmic string parameter $\alpha$. Similarly, on the left side, we have $b = 2.3\,M>2\,M$ and $a = 1.5\,M$, and on the right side, we have $b = 2.3\,M$ and $a = 2\,M$. Here as well, we observe that the effective potential increases as the value of the cosmic string parameter decreases. Furthermore, in Figures 3 and 4, we depict the effective potential for time-like geodesics, which correspond to massive objects. On the left side, $b = 2.3\,M>2\,M$ and $a = 1.5\,M$, and on the right side, we have $b = 2.3\,M>2\,M$ and $a = 2\,M$. We observe that the same effects seen in the case of null geodesics apply here as well, with the presence of the cosmic string leading to an increased effective potential.

Now, we discuss circular photon orbits which satisfies the conditions $\dot{r}=0$ and $\ddot{r}=0$. Unstable photon orbits, which constitute the so-called photon sphere, additionally satisfy $\dddot{r}>0$. Generally, for unstable photon orbit (photon sphere), the following conditions in terms of the effective potential satisfies
\begin{equation}
    \frac{dV_{eff}}{dr}|_{r=r_{ph}}=0,\quad \frac{d^2V_{eff}}{dr^2}|_{r=r_{ph}}<0,
    \label{photon}
\end{equation}
where $r_{ph}$ is the radius of a photon sphere. Note that a photon sphere corresponds to the maximum potential. For the space-time (\ref{a1}), using the first condition, we obtain 
\begin{equation}
    r^{3}_{ph}-M\,r^{2}_{ph}-2\,M\,r_{ph}+M\,(b^2-a^2)=0
    \label{photon2}
\end{equation}
whose real valued solution gives us the radius of the photon sphere that exists for $b^2 \geq a^2$ and $(b^2-a^2) < 4\,M^2$ with $2\,M < b < 3\,M$, and $a \geq 0$. Thus, the exact location of the photon sphere depends on the parameters $a, b$ and $M$ which is not located at the wormhole throat. For the case $b^2=a^2$, from above equation (\ref{photon2}), we find this photon sphere radius $r_{ph}=\frac{M}{2}+\sqrt{M\,\Big(\frac{M}{4}+2\Big)}$ that solely depends on mass of the object $M$.

To find the path equation for photon ray, from Eqs. (\ref{5}) and (\ref{6}), we obtain
\begin{equation}
    \Big(\frac{dr}{d\tau}\Big)^2=E^2-\Big(1-\frac{2\,M}{r}\Big)\,\frac{J^2/\alpha^2}{(r^2-f^2)}.
    \label{ph1}
\end{equation}
Transforming $r=\frac{1}{u(\phi)}$ into the above equation and then differentiating w. r. t. $\phi$, one will find the following path equation by neglecting the terms $\alpha^4$ (assuming $\alpha$ has positive finite value but not so small)
\begin{equation}
    \frac{1}{\alpha^2}\,\Big(\frac{d^2u}{d\phi^2}\Big)+u=3\,M\,u^2+2\,(b^2-a^2)\,u^3-5\,M\,(b^2-a^2)\,u^4.
    \label{ph2}
\end{equation}
For the case $b^2=a^2$ and $\alpha \to 1$, one will find similar path equation for photon ray as in the case of Schwarzschild solution \cite{SW}. Thus, we see that the presence of various parameters $a, b$ and the cosmic string parameter $\alpha$ influences the path equation of photon ray.

Overall, our analysis has demonstrated that the effective potential of the system and the path equation of photon ray in the backdrop of Schwarzschild-Klinkhamer wormhole is influenced by the parameters $a, b$ and the cosmic string characterized by the parameter $\alpha$. The effective potential increases with the presence of cosmic string, as confirmed by our plots for both null and time-like geodesics. 

\subsection{Gravitational lensing by wormhole with a cosmic string for M=0}

In this part, we focus on analyzing the deflection angle of photon light within the space-time geometry given by (\ref{a1}). Specifically, we consider here the case where mass of the object is zero, $M=0$. By evaluating Eqs. (\ref{4}) and (\ref{5}) for null geodesics, we derive the following expression:  
\begin{eqnarray}
    \frac{d\phi}{dr}=\frac{\dot{\phi}}{\dot{r}}&=&\frac{J}{\alpha^2}\,\frac{1}{(r^2-f^2)\,\sqrt{E^2-2\,V_{eff}(r)}}=\frac{\beta}{\alpha^2}\,\frac{1}{\sqrt{(r^2-f^2)^2-\frac{\beta^2}{\alpha^2}\,(r^2-f^2)}}
    \label{8}
\end{eqnarray}
where $\beta=\frac{J}{E}$ is the impact parameter characterizing a particular null geodesic with the conserved energy parameter $E$ and the angular momentum $J$, and we have defined $f^2=(b^2-a^2)\geq 0$ for $b^2 \geq a^2$. 

Hence, we can obtain the deflection angle \cite{aa6,aa8,aa28,aa31,hh1,aa33} given by
\begin{eqnarray}
\delta\phi=\Delta\phi-\pi,
    \label{9}
\end{eqnarray}
where
\begin{eqnarray}
    \Delta\phi=\frac{2\,\beta}{\alpha^2}\,\int^{\infty}_{r_0}\,\frac{dr}{\sqrt{(r^2-f^2)^2-\frac{\beta^2}{\alpha^2}\,(r^2-f^2)}}
    =\frac{2\,\beta}{\alpha^2}\,\int^{\infty}_{r_0}\,\frac{dr}{\sqrt{(r^2-f^2)(r^2-r^2_{0})}}.
    \label{10}
\end{eqnarray}
Here $r=r_0$ is the turning point which can be obtained by setting $\Big(\frac{dr}{d\phi}\Big)|_{r=r_0}=0$ for null geodesics, and in terms of the impact parameter $\beta$, we obtain
\begin{equation}
     r_0=\sqrt{f^2+\frac{\beta^2}{\alpha^2}}=\sqrt{b^2-a^2+\frac{\beta^2}{\alpha^2}}.
    \label{11}
\end{equation}

Defining a new coordinate via $y=r_0/r$ and $g=f/r_0$ into the integral (\ref{10}), we obtain
\begin{eqnarray}
    \Delta\phi=\frac{2\,\beta}{\alpha^2\,r_0}\,K(g),
    \label{12}
\end{eqnarray}
where the complete elliptic integral of the first kind is given by \cite{MA}
\begin{equation}
    K(g)=\int^{1}_{0}\,\frac{dy}{\sqrt{(1-g^2\,y^2)(1-y^2)}},\quad 0 < g <1.
    \label{13}
\end{equation}
The expression of this complete elliptic integral will be  
\begin{eqnarray}
K(g)=\frac{\pi}{2}\,\sum^{\infty}_{n=0}\,\Bigg(\frac{(2\,n)!}{2^{2\,n}\,(n!)^2} \Bigg)^2\,g^{2\,n}=\frac{\pi}{2}\,\Big[1+\frac{g^2}{4}+\frac{9\,g^4}{64}+\frac{2\,5}{256}\,g^6+...\Big],
    \label{14}
\end{eqnarray}
for $g<1$ since $f^2 < r^{2}_0$, that is, in the weak field limit.

Therefore, the deflection angle from (\ref{9}) using (\ref{12}) becomes
\begin{eqnarray}
    \delta\phi (g)=\frac{2\,\beta}{\alpha^2\,\sqrt{b^2-a^2+\frac{\beta^2}{\alpha^2}}}\,K(g)-\pi.
    \label{15}
\end{eqnarray}
We illustrate the deflection angle given by equation (\ref{15}) as a function of $g$ in Figure 5, considering both cases: without the effects of cosmic string and with the effects of cosmic string. It is observed that the deflection angle is significantly increased in the presence of cosmic string, indicating the substantial influence of the cosmic string on the deflection of photon light.  

\begin{figure}
    \includegraphics[width=0.45\textwidth]{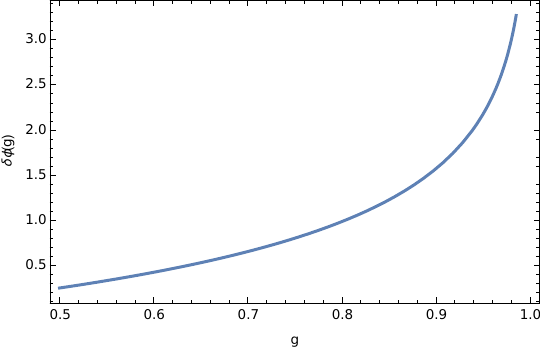}
    \hfill
    \includegraphics[width=0.45\textwidth]{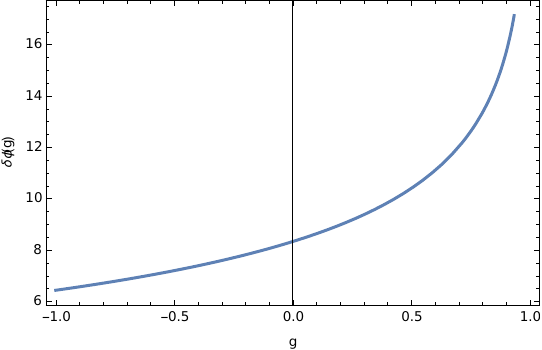}
    \caption{\label{fig: 5} Deflection angle without ($\alpha=1$, left one) and with ($\alpha=1/2$, right one) cosmic string parameter. Here, $a=2.2$, $b=2.3$, $\beta=1.5$.}
\end{figure}

As the parameter $g$ always falls within the range $0 < g < 1$, in the weak field limit, we can express the deflection angle equation (\ref{15}) as follows:
\begin{eqnarray}
\delta\phi=\frac{\pi\,\beta}{\alpha^2\,\sqrt{b^2-a^2+\frac{\beta^2}{\alpha^2}}}\,\Bigg[1+\frac{(b^2-a^2)/4}{\Big(b^2-a^2+\frac{\beta^2}{\alpha^2}\Big)}+\frac{9(b^2-a^2)^2/64}{\Big(b^2-a^2+\frac{\beta^2}{\alpha^2}\Big)^2}+\mathcal{O} (g)^6\Bigg]-\pi.
    \label{16}
\end{eqnarray}

It is evident that the deflection angle of photon light is influenced by the cosmic string characterized by the parameter $\alpha$ and undergoes modifications. When $b^2=a^2$, the space-time corresponds to a Schwarzschild-Klinkhamer wormhole solution with cosmic string, as given by the metric (\ref{a7}). In that case, the expression of the deflection angle for null geodesics in the weak field limit, derived from equation (\ref{16}), can be written as $\delta\phi=\pi\,\left(\frac{1}{\alpha}-1\right)>0$.

\subsection{Gravitational lensing by wormhole with a cosmic string for $M \neq 0$}

In this part, we study the deflection of photon ray for nonzero mass, $M \neq 0$. Therefore, from equation (\ref{5}) using (\ref{6}) for null geodesics, we can write
\begin{equation}
    \dot{r}^2=\Big(\frac{dr}{d\tau}\Big)^2=E^2-\Big(1-\frac{2\,M}{r}\Big)\,\frac{J^2/\alpha^2}{(r^2-f^2)}.
    \label{17}
\end{equation}
From (\ref{4}) and (\ref{17}), we derive the following expression 
\begin{eqnarray}
    \frac{dr}{d\phi}=\frac{\alpha^2}{\beta}\,\sqrt{(r^2-f^2)-\frac{\beta^2}{\alpha^2}\,\Big(1-\frac{2\,M}{r}\Big)\,(r^2-f^2)}.
    \label{18}
\end{eqnarray}
Therefore, the deflection angle of photon ray is given by
\begin{equation}
    \delta\phi=\Delta\phi-\pi,
    \label{19}
\end{equation}
where
\begin{equation}
    \Delta\phi=\frac{2}{\alpha}\,\int^{\infty}_{\bar{r}}\,\frac{dr}{\sqrt{r^2-f^2}\,\sqrt{\alpha^2\,\Big(\frac{r^2-f^2}{\beta^2}\Big)-\Big(1-\frac{2\,M}{r}\Big)}}.
    \label{20}
\end{equation}
Here $r=\bar{r}$ is the turning point and can be obtained by setting $\frac{dr}{d\phi}|_{r=\bar{r}}=0$ for null geodesics, and one will find impact parameter in terms of this turning point by $\beta (\bar{r})=\alpha\,\sqrt{\bar{r}^2-f^2}\,\Big(1-\frac{2\,M}{\bar{r}}\Big)^{-1/2}$. Thus, equation (\ref{20}) can be expressed by replacing the impact parameter as follows: 
\begin{equation}
    \Delta\phi=\frac{2}{\alpha}\,\int^{\infty}_{\bar{r}}\,\frac{dr}{\sqrt{r^2-f^2}\,\sqrt{\Big(\frac{r^2-f^2}{\bar{r}^2-f^2}\Big)\Big(1-\frac{2\,M}{\bar{r}}\Big)-\Big(1-\frac{2\,M}{r}\Big)}}.
    \label{21}
\end{equation}

If one consider $b^2=a^2$, as discussed in {\bf CASE B} in {\it section 2}, space-time (\ref{a1}) under this condition becomes a vacuum Schwarzschild-Klikhamer wormhole with a cosmic string. Therefore, for $b^2=a^2$, we have found $f \to 0$, and thus, the impact parameter in terms of the turning point will be $\beta (\bar{r}_0)=\alpha\,\bar{r}_0\,\Big(1-\frac{2\,M}{\bar{r}_0}\Big)^{-1/2}$, where $\bar{r}_0>2\,M$. In that case, from equation (\ref{21}), we obtain
\begin{equation}
    \Delta\phi=\frac{2}{\alpha}\,\int^{\infty}_{\bar{r}_0}\,\frac{dr}{r\,\sqrt{\Big(\frac{r}{\bar{r}_0}\Big)^2\Big(1-\frac{2\,M}{\bar{r}_0}\Big)-\Big(1-\frac{2\,M}{r}\Big)}}=\frac{1}{\alpha}\,\Delta\phi_{Sch}
    \label{22}
\end{equation}
which is equals to $\frac{1}{\alpha}$ times a similar expression obtained in the background of Schwarzschild solution \cite{SW}. Here, $r$ coordinate lies in the interval $b \leq r < \infty$, where $ 2\,M < b <3\,M$ instead of $r \in[0, \infty)$. We see that the presence of cosmic string in the space-time geometry increases the deflection angle compared to the known expression in Schwarzschild solution.

\section{Schwarzschild-Klinkhamer wormhole with a global monopole}

In this section, we consider a generalized version of the Klinkhamer wormhole space-time (\ref{K1}) by incorporating a factor $A(x)$ analogue to the Schwarzschild solution in the presence of a global monopole. Therefore, the line-element describing this wormhole with a global monopole is given by
\begin{eqnarray}
      ds^2=-A(x)\,dt^2+\frac{dx^2}{\alpha^{2}_0\,\Big(1+\frac{b^2}{x^2}\Big)\,A(x)}+(x^2+a^2)(d\theta^2+\sin^2 \theta\,d\phi^2),\, A(x)=\Big[1-\frac{2\,M}{\sqrt{x^2+b^2}}\Big],
\label{aa1}
\end{eqnarray}
where $0 < \alpha_0 <1$ is the global monopole parameter. The introduction of a global monople into the space-time brings about significant alterations to its geometry and curvature characteristics. This wormhole space-time (\ref{aa1}) possesses the following nonzero components of the Ricci tensor $R_{\mu\nu}$ given by
\begin{eqnarray}
    &&R_{tt}=\frac{2\,M\,\alpha^2_{0}\,(b^2-a^2)}{(x^2+a^2)\,(x^2+b^2)^{3/2}}\,\Big(1-\frac{2\,M}{\sqrt{x^2+b^2}}\Big),\nonumber\\
    &&R_{xx}=\frac{2\,M\,x^2}{(x^2+b^2)^{5/2}}\,\Big(1-\frac{2\,M}{\sqrt{x^2+b^2}}\Big)^{-1}+\frac{2}{(x^2+a^2)^2}\,\Bigg[-a^2+\frac{x^2+a^2}{x^2+b^2}\,\Bigg\{b^2+\frac{M\,x^2}{2\,M-\sqrt{x^2+b^2}}\Bigg\} \Bigg],\nonumber\\
    &&R_{\theta\theta}=(1-\alpha^2_{0}),\quad R_{\phi\phi}=(1-\alpha^2_{0})\,\sin^2 \theta.
    \label{riccitensor}
\end{eqnarray}
The Ricci $R$ and Kretschmann scalar curvatures $\mathcal{K}$ for the metric (\ref{aa1}) are given by
\begin{eqnarray}
    &&R=2\,\Bigg[\frac{1}{x^2+a^2}+\alpha^2_{0}\,\Bigg\{\frac{2\,M}{(x^2+b^2)^{3/2}}+\frac{-2\,a^2+b^2-x^2-2\,M\,\sqrt{x^2+b^2}}{(x^2+a^2)^2}\Bigg\}\Bigg],\nonumber\\
    &&\mathcal{K}=\frac{4}{(x^2+a^2)^4\,(x^2+b^2)^3}\,\Big[4\,M\,(a^2 + x^2)\,(b^2 + x^2)^{7/2}\,\alpha^{2}_{0}+4\,M^2\,(a^2 + x^2)^4\,\alpha^{4}_{0}\nonumber\\
    &&-4\,M\, (a^2 + x^2)^2\,(b^2 + x^2)^{5/2}\,\alpha^{4}_{0}-12\,b^2\,M\,(b^2 + x^2)^{7/2}\,\alpha^{4}_{0}-12\,M\,x^2\,(b^2+x^2)^{7/2}\,\alpha^{4}_{0}\nonumber\\
    &&+12\,M\,(a^2 + x^2)\,(b^2 + x^2)^{7/2}\,\alpha^{4}_{0}+3\,(b^2 + x^2)^4\,(b^2 + 4 M^2 + x^2)\,\alpha^{4}_{0}\nonumber\\
    &&-2\,(a^2 + x^2)\,(b^2+ x^2)^3\,\alpha^{2}_{0}\,\Big(b^2 + x^2+2\,(b^2 + 2 M^2 + x^2)\,\alpha^{2}_{0}\Big)\nonumber\\
    &&+(a^2 + x^2)^2\,(b^2 +x^2)^2\,\Big(b^2 + x^2 + 2\,(b^2 + 2 M^2 + x^2)\,\alpha^{4}_{0}\Big)\Big],
    \label{riccitensor2}
\end{eqnarray}
which are finite at $x=0$ even for the case $b^2=a^2$ and vanishes for $x \pm\,\infty$. 

Considering the energy-momentum tensor $T^{\mu}_{\nu}=\mbox{diag}(-\rho, p_x, p_t, p_t)$, where symbols have their usual meanings, we obtain 
\begin{eqnarray}
    &&\rho=-T^{t}_{t}=-G^{t}_{t}=\frac{2\,M\,\alpha^2_{0}}{(x^2+a^2)\,\sqrt{x^2+b^2}}+\frac{1}{x^2+a^2}  +\frac{\alpha^2_{0}\Big(-2\,a^2+b^2-x^2-2\,M\,\sqrt{x^2+b^2}\Big)}{(x^2+a^2)^2},\nonumber\\
    &&p_x=T^{x}_{x}=G^{x}_{x}=\frac{2\,\alpha^2_{0}\,(1+b^2/x^2)}{(x^2+a^2)^2}\,\Big(1-\frac{2\,M}{\sqrt{x^2+b^2}}\Big)\,\Bigg[-a^2+\Big(\frac{x^2+a^2}{x^2+b^2}\Big)\,\Bigg\{b^2+\frac{M\,x^2}{2\,M-\sqrt{x^2+b^2}}\Bigg\}\Bigg]\nonumber\\
    &&-\frac{1}{x^2+a^2}
    -\frac{\alpha^2_{0}\,\Big(-2\,a^2+b^2-x^2-2\,M\,\sqrt{x^2+b^2}\Big)}{(x^2+a^2)^2},\nonumber\\
    &&p_{t}=T^{\theta}_{\theta}=G^{\theta}_{\theta}=G^{\phi}_{\phi}=-\alpha^2_{0}\,\Bigg[\frac{2\,M}{(x^2+b^2)^{3/2}}+\frac{b^2-a^2-2\,M\,\sqrt{x^2+b^2}}{(x^2+a^2)^2}\Bigg].
    \label{energy-density}
\end{eqnarray}

From above we see that the wormhole space-time considered by (\ref{aa1}) is a non-vacuum solution of the field equations. The energy-density given by $\rho=\Bigg[\frac{2\,M\,\alpha^2_{0}\,(b^2-a^2)}{(x^2+a^2)\,(x^2+b^2)^{3/2}}+\frac{R}{2}\Bigg]>0$ is positive provided $b^2 \geq a^2$, where $b>2\,M$. All these physical quantities, such as the energy-density ($\rho$), the tangential pressure ($p_t$), and the scalar curvatures given by (\ref{riccitensor2}) are finite at $x=0$ and vanishes for $x \pm\,\infty$. Note that the pressure component $p_x$ is undefined at $x=0$ under the case $b^2 >a^2$. However, for $b^2=a^2$ in the subsequent discussion, we show that this pressure component is finite at $x=0$ and vanishes for $x \pm\,\infty$.

Now, we discuss below a few special cases of the above Schwarzschild-Klinkhamer wormhole metric with a global monopole.

\vspace{0.2cm}

{\bf CASE A}: If one choose $b^2=a^2$, then the space-time (\ref{aa1}) reduces to the following form 
\begin{eqnarray}
    &&ds^2=-\Big(1-\frac{2\,M}{\sqrt{x^2+b^2}}\Big)\,dt^2+\Bigg[\alpha^{2}_0\,\Bigg(1-\frac{2\,M}{\sqrt{x^2+b^2}}\Bigg)\,\Bigg(1+\frac{b^2}{x^2}\Bigg)\Bigg]^{-1}dx^2\nonumber\\
    &&+(x^2+b^2)(d\theta^2+\sin^2 \theta\,d\phi^2).
    \label{aa2}
\end{eqnarray}

This space-time (\ref{aa2}) is a new non-vacuum defect wormhole solution with a global monopole and is a degenerate one. This wormhole space-time possesses the following nonzero components of the Einstein tensor, $G^{\mu}_{\,\nu}$ given by
\begin{equation}
    G^{t}_{t}=G^{x}_{x}=-\frac{(1-\alpha^2_{0})}{x^2+b^2}=-R/2
    \label{aa3}
\end{equation}
and $G^{\theta}_{\,\theta}=0=G^{\phi}_{\,\phi}$.

The Ricci scalar $R$ and the Kretschmann scalar curvature $\mathcal{K}$ of this new wormhole (\ref{aa2}) are given by 
\begin{equation}
    R=\frac{2\,(1-\alpha^2_{0})}{x^2+b^2}\quad,\quad \mathcal{K}=\frac{2\,\Big[289\,(x^2+b^2)-1248\,M\,\sqrt{x^2+b^2}+2124\,M^2\Big]}{(x^2+b^2)^{3/2}}.
    \label{aa4}
\end{equation}
One can see from above that both the scalar curvatures are finite at $x=0$ and vanishes for $x \to \pm\,\infty$. Assuming that this new non-vacuum wormhole (\ref{aa2}) is a solution of the Einstein's field equations, $G^{\mu}_{\,\nu}=T^{\mu}_{\,\nu}$, where $\frac{8\,\pi\,G}{c^4}=1$, then energy-density of the stress-energy fluid is given by
\begin{equation}
\rho=-T^{t}_{t}=-G^{t}_{t}=R/2=\frac{1-\alpha^2_{0}}{x^2+b^2}>0
\label{energy}
\end{equation}
which is always positive since global monopole parameter lies $0 < \alpha_0 <1$ in the gravitation and cosmology. The only pressure component along $x$-direction is given by 
\begin{equation}
p_{x}=-R/2=\frac{\alpha^2_{0}-1}{x^2+b^2}<0
\label{energy2}
\end{equation}
with the tangential component $p_{t}=0$. One can see that $\rho+p_{x}=0$ which implies that the null energy condition (NEC) is automatically satisfied.

To check the validate of the energy conditions, we consider the following vector fields, such as the null vector $k^{\mu}$, the time-like four-vector $U^{\mu}$ and a spacelike unit vector $\eta^{\mu}$ along $x$-direction for the space-time (\ref{aa2}) defined by
\begin{eqnarray}
    &&k^{\mu}=\frac{1}{\sqrt{2}}\,\Bigg[-\Big(1-\frac{2\,M}{\sqrt{x^2+b^2}}\Big)^{-1/2}, \alpha_0\,\sqrt{1+\frac{b^2}{x^2}}\,\Big(1-\frac{2\,M}{\sqrt{x^2+b^2}}\Big)^{1/2}, 0, 0\Bigg],\nonumber\\
    &&U^{\mu}=\Big(1-\frac{2\,M}{\sqrt{x^2+b^2}}\Big)^{-1/2}\,\delta^{\mu}_{0},\nonumber\\
    &&\eta^{\mu}=\Bigg[\alpha_0\,\Big(1-\frac{2\,M}{\sqrt{x^2+b^2}}\Big)^{1/2}\,\sqrt{1+\frac{b^2}{x^2}}\Bigg]\,\delta^{\mu}_{x},
    \label{null}
\end{eqnarray}
where the vector fields satisfy the following relations
\begin{equation}
    -U^{\mu}\,U_{\mu}=1=\eta^{\mu}\,\eta_{\mu},\quad U_{\mu}\,\eta^{\mu}=0=k^{\mu}\,k_{\mu},\quad k_{\mu}\,\eta^{\mu}=\frac{1}{\sqrt{2}}=U_{\mu}\,k^{\mu}.\label{null1}
\end{equation}

For zero tangential pressure, $p_{t}=0$, we choose the following energy-momentum tensor 
\begin{eqnarray}
    T_{\mu\nu}=(\rho+p_{t})\,U_{\mu}\,U_{\nu}+p_{t}\,g_{\mu\nu}+(p_{x}-p_{t})\,\eta_{\mu}\,\eta_{\nu}=\rho\,U_{\mu}\,U_{\nu}+p_{x}\,\eta_{\mu}\,\eta_{\nu}.
    \label{null2}
\end{eqnarray}
With the help of (\ref{null}), we have the following energy conditions
\begin{equation}
    T_{\mu\nu}\,U^{\mu}\,U^{\nu}=\rho>0\quad \mbox{and} \quad T_{\mu\nu}\,k^{\mu}\,k^{\nu}=0.
    \label{null3}
\end{equation}

Thus, we see that the new wormhole metric (\ref{aa2}) is a solution of the field equations satisfying the weak energy condition (WEC) and the null energy condition (NEC). We called this new non-vacuum wormhole (\ref{aa2}) as Schwarzschild-Klinkhamer (SK) defect wormhole metric with a global monopole. Since the space-time described by the line-element (\ref{aa1}) is a generalization of the non-vacuum space-time (\ref{aa2}) under the case $b^2 \neq a^2$, we called this new metric (\ref{aa1}) the generalized SK-wormhole space-time with a global monopole which satisfies both the weak energy condition (WEC) and the null energy condition (NEC).

If one perform a transformation via $r=\sqrt{x^2+b^2}$ into the space-time (\ref{aa2}), we obtain the following line-element 
\begin{eqnarray}
    ds^2=-\Big(1-\frac{2\,M}{r}\Big)\,dt^2+\alpha^{-2}_0\,\Big(1-\frac{2\,M}{r}\Big)^{-1}\,dr^2+r^2\,(d\theta^2+\sin^2 \theta\,d\phi^2),\, r \in [b>2\,M, \infty)
    \label{aa5}
\end{eqnarray}
which looks similar form to the Schwarzschild space-time with a global monopole but it is a non-vacuum solution of the field equations.

\vspace{0.2cm}

{\bf CASE B}: In the limit $M=0$, space-time (\ref{aa1}) reduces to the following form 
\begin{equation}
      ds^2=-dt^2+\Big(1+\frac{b^2}{x^2}\Big)^{-1}\,\alpha^{-2}_0\,dx^2+(x^2+a^2)\,(d\theta^2+\sin^2 \theta\,d\phi^2)
    \label{aa6}
\end{equation}
the Klinkhamer defect-wormhole space-time with a global monopole \cite{ss1}, which is a non-vacuum solution of the field equations. The Ricci scalar and the Kretschamnn scalar curvatures are given by
\begin{eqnarray}
    &&R=\frac{2\,[x^2+a^2+(-2\,a^2+b^2-x^2)\alpha^2_{0}]}{(x^2+a^2)^2},\nonumber\\
    &&\mathcal{K}=\frac{2\,[371\,a^4+706\,b^4+624\,b^2\,x^2+289\,x^4- 
   2\,a^2\,(394\,b^2 + 23\,x^2)]}{(a^2 + x^2)^4}.
   \label{aa7}
\end{eqnarray}
Both the scalar curvatures are finite at $x=0$ and vanishes for $x \to \pm\,\infty$. 

Even for $b^2=a^2$, metric (\ref{aa6}) is still a non-vacuum solution of the field equations with the energy-density $\rho=R/2$, and the radial pressure $p_{r}=-R/2<0$, where the Ricci scalar $R$ is given by (\ref{aa4}), and the Kretschmann scalar be $\mathcal{K}=\frac{578}{(x^2+b^2)}$. All these physical quantities are finite at $x=0$ and vanishes for $x \to \pm\,\infty$. In the limit $\alpha_0 \to 1$, space-time (\ref{aa6}) reduces to the original Klinkhamer wormhole (see Eq. (4) in \cite{ss1}), and becomes a vacuum-defect one \cite{sss1} provided for $b^2=a^2$. 

For the case $b^2=a^2$, one can do a transformation via $r=\sqrt{x^2+b^2}$ into the space-time (\ref{aa6}), we obtain
\begin{equation}
    ds^2=-dt^2+\frac{dr^2}{\alpha^2_{0}}+r^2\,(d\theta^2+\sin^2 \theta\,d\phi^2),\quad r \in [b>0, \infty)
    \label{aa12}
\end{equation}
which looks similar form to a global monopole-like space-time \cite{ss7,SR}.

\vspace{0.2cm}

{\bf CASE C}: In the limit $M \to 0$, and $a=0$, space-time (\ref{aa1}) reduces to the following form
\begin{equation}
      ds^2=-dt^2+\alpha^{-2}_0\,\Big(1+\frac{b^2}{x^2}\Big)^{-1}\,dx^2+x^2\,(d\theta^2+\sin^2 \theta\,d\phi^2)
    \label{aa8}
\end{equation}
an example of a topologically charged Eddington-inspired Born-Infeld (EiBI) gravity background space-time. The Ricci scalar and the Kretschmann scalar curvatures given by 
\begin{equation}
    R=\frac{2\,[(1-\alpha^2_{0})\,x^2+b^2\,\alpha^2_{0}]}{x^4} \quad,\quad \mathcal{K}=\frac{2\,[706\,b^4+624\,b^2\,x^2+289\,x^4]}{x^8}
    \label{aa9}
\end{equation}
vanishes at $x \to \pm\,\infty$. Recently, examples of a topologically charged Eddington-inspired Born-Infeld gravity background space-time has been discussed in Refs. \cite{aa33,ff}. 

\vspace{0.2cm}

{\bf CASE D}: In the limit $M \to 0$, and $b=0$, space-time (\ref{aa1}) reduces to the following form
\begin{equation}
      ds^2=-dt^2+\frac{dx^2}{\alpha^2_{0}}+(x^2+a^2)\,(d\theta^2+\sin^2 \theta\,d\phi^2)
    \label{aa10}
\end{equation}
a topologically charged Ellis-Bronnikov-type wormhole \cite{sss}, and becomes an Ellis-Bronnikov-type or Morris-Thorne-type wormhole for $\alpha_0 \to 1$ \cite{HGE,KAB,MT}.

\section{Geodesics motion of test particles by wormhole with a global monopole}

In this part, we study the geodesics motions of test particles in wormhole (\ref{aa1}) background. Therefore, the Lagrangian of the system on the equatorial plane $\theta=\frac{\pi}{2}$ is given by
\begin{equation}
\mathcal{L}=\frac{1}{2}\,\Bigg[-A(x)\,\dot{t}^2+\frac{1}{\alpha^2_{0}\,\Big(1+\frac{b^2}{x^2}\Big)}\,\frac{\dot{x}^2}{A(x)}+(x^2+a^2)\,\dot{\phi}^2\Bigg].
    \label{f1}
\end{equation}
Here also, two constant of motions corresponding to the coordinate $(t, \phi)$ which are as follows: 
\begin{eqnarray}
    A(x)\,\dot{t}=E\Rightarrow \dot{t}=\frac{E}{A(x)}\quad \mbox{and} \quad (x^2+a^2)\,\dot{\phi}=J_0\Rightarrow \dot{\phi}=\frac{J_0}{(x^2+a^2)},
    \label{f3}
\end{eqnarray}
where $E$ is the conserved energy parameter, and $J_0$ is the conserved angular momentum. 

Therefore, the Lagrangian (\ref{f1}) for null or time-like geodesics after transforming $r=\sqrt{x^2+b^2}$ becomes
\begin{equation}
    \frac{1}{2}\,\Big(\frac{dr}{d\tau}\Big)^2+V_{eff} (r)=\frac{\alpha^2_{0}\,E^2}{2}=\frac{E^2_{eff}}{2},
    \label{f4}
\end{equation}
which has the same form of the equation for a unit mass particle with energy $E^2_{eff}/2$ moving in one-dimensional effective potential given by
\begin{equation}
    V_{eff} (r)=\alpha^2_{0}\,\Big(\frac{1}{2}-\frac{M}{r}\Big)\,\Bigg[\frac{J^2_{0}}{r^2+a^2-b^2}-\varepsilon\Bigg],
    \label{f5}
\end{equation}
where $\varepsilon=0$ for null geodesics and $\varepsilon=-1$ for time-like geodesics.

We see that the effective potential (\ref{f6}) for null or time-like geodesics is influenced not only by the parameters $a, b$, and $M$ but also by the global monopole characterized by the parameter $\alpha_0$. One can see that the effective potential of the system decreases compared to the result without the effects of global monopole, since in the context of gravitation and cosmology, the values of the global monopole parameter $\alpha_0$ typically lies $0 < \alpha_0 < 1$.

In Figure 6, we present plots of the effective potential for null geodesics without and with the effects of the global monopole parameter: $\alpha_0 = 1$ and $\alpha_0 = 0.5$. On the left side, we set $b = 2.3\,M>2\,M$ and $a = 1.5\,M$, while on the right side, we use $b = 2.3\,M>2\,M$ and $a = 2\,M$. It can be observed that the effective potential for null geodesics is diminished under the influence of the global monopole, as compared to the case without the presence of global monopole parameter ($\alpha_0 \to 1$). 

Moving on to Figure 7, we plot the effective potential for null geodesics considering various values of the global monopole parameter $\alpha_0$. Similarly, on the left side, we have $b = 2.3\,M>2\,M$ and $a = 1.5\,M$, and on the right side, we have $b = 2.3\,M>2\,M$ and $a = 2\,M$. Here as well, we observe that the effective potential increases as the value of the global monopole parameter $\alpha_0$ increases.

To find the path equation for photon ray, from Eqs. (\ref{f4}) and (\ref{f5}), we obtain
\begin{equation}
    \Big(\frac{dr}{d\tau}\Big)^2=\alpha^2_{0}\,\Bigg[E^2-\Big(1-\frac{2\,M}{r}\Big)\,\frac{J_{0}^2}{(r^2-f^2)}\Bigg].
    \label{pph1}
\end{equation}
Transforming $r=\frac{1}{u(\phi)}$ into the above equation and then differentiating w. r. t. $\phi$, one will find the following path equation in terms of the impact parameter $\beta'(=J_0/E)$ given by
\begin{equation}
    \frac{1}{\alpha^2_{0}}\,\Big(\frac{d^2u}{d\phi^2}\Big)+\Bigg[1+\frac{2\,(b^2-a^2)}{\beta'^2}\Bigg]\,u=3\,M\,u^2+2\,(b^2-a^2)\,\Bigg[1+\frac{b^2-a^2}{\beta'^2}\Bigg]\,u^3-5\,M\,(b^2-a^2)\,u^4.
    \label{pph2}
\end{equation}

In summary, plots of the effective potential for null geodesics and the path equation of photon light have demonstrated the significant influence of global monopole characterized by the parameter $\alpha_0$. The effective potential decreases since $\alpha_0 <1$ indicating the impact of global monopole on the dynamics of test particles moving along the geodesics in this system.

\begin{figure}[tbp]
    \includegraphics[width=0.45\textwidth]{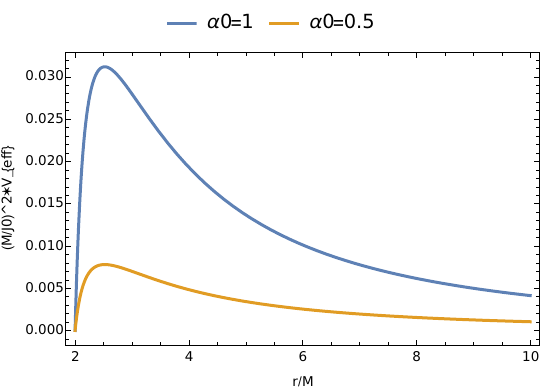}
    \hfill
    \includegraphics[width=0.45\textwidth]{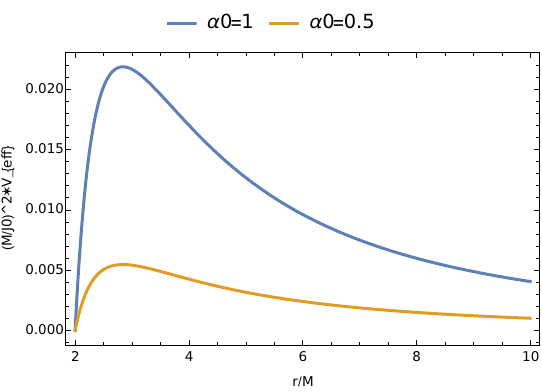}
    \caption{\label{fig: 6} Effective potential for light-like geodesics without ($\alpha_0=1$) and with global monopole ($\alpha_0=0.5$) parameter. Here, $a=1.5\,M$, $b=2.3\,M$ (left); $a=2\,M$, $b=2.3\,M$ (right)}
\end{figure}

\begin{figure}[tbp]
    \includegraphics[width=0.45\textwidth]{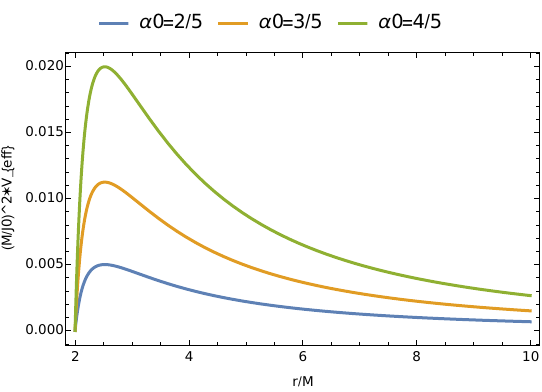}
    \hfill
    \includegraphics[width=0.45\textwidth]{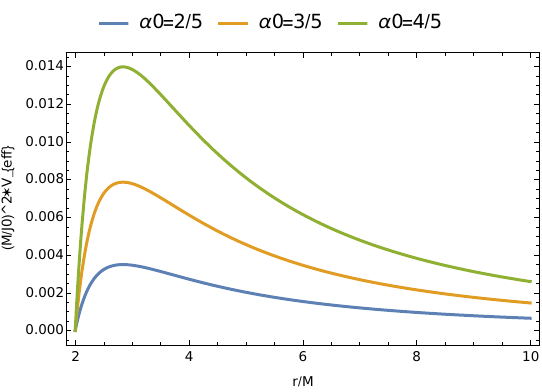}
    \caption{\label{fig: 7} Effective potential for light-like geodesics for different values of global monopole parameter $\alpha_0$. Here, $a=1.5\,M$, $b=2.3\,M$ (left); $a=2\,M$, $b=2.3\,M$ (right) }
\end{figure}

\subsection{Gravitational lensing by wormhole with a global monopole for M =0}

In this part, we calculate the deflection angle of photon rays within the space-time geometry described by equation (\ref{aa1}) for the specific case where $M=0$. By considering Eqs. (\ref{f3}) and (\ref{f4}) for the null geodesics with $M=0$, we obtain the following expression:   
\begin{eqnarray}
    \frac{d\phi}{dr}=\frac{\dot{\phi}}{\dot{r}}=\frac{\beta'}{\alpha_0}\,\frac{1}{\sqrt{(r^2-f^2)^2-\beta'^2\,(r^2-f^2)}}.
    \label{f6}
\end{eqnarray}
Hence, the deflection angle is given by \cite{aa28}
\begin{eqnarray}
\delta\phi=\Delta\phi-\pi,
    \label{f7}
\end{eqnarray}
where
\begin{eqnarray}
    \Delta\phi=\frac{2\,\beta'}{\alpha_0}\,\int^{\infty}_{r_0}\,\frac{dr}{\sqrt{(r^2-f^2)(r^2-r^2_{0})}},
    \label{f8}
\end{eqnarray}
where $r=r_0$ is the turning point obtained by setting $\Big(\frac{dr}{d\phi}\Big)=0$ in equation (\ref{f4}) for the null geodesics corresponding to the case $M=0$, and in terms of impact parameter $\beta'$, it is given by
\begin{equation}
    r_0=\sqrt{f^2+\beta'^2}=\sqrt{b^2-a^2+\beta'^2}.
    \label{f9}
\end{equation}
Defining $z=r_0/r$ and $g'=f/r_0$ into the integral (\ref{f8}), we obtain
\begin{eqnarray}
    \Delta\phi=\frac{2\,\beta'}{\alpha_0\,r_0}\,K(g'),
    \label{f10}
\end{eqnarray}
where the complete elliptic integral of the first kind is given by \cite{MA}
\begin{equation}
    K(g')=\int^{1}_{0}\,\frac{dy}{\sqrt{(1-g'^2\,z^2)(1-z^2)}},\quad 0 < g' <1.
    \label{f11}
\end{equation}
The expression of this complete elliptic integral will be 
\begin{eqnarray}
    K(g')=\frac{\pi}{2}\,\sum^{\infty}_{n=0}\,\Bigg(\frac{(2\,n)!}{2^{2\,n}\,(n!)^2} \Bigg)^2\,g'^{2\,n}=\frac{\pi}{2}\,\Big[1+\frac{g'^2}{4}+\frac{9\,g'^4}{64}+\frac{2\,5}{256}\,g'^6+......\Big]
    \label{f12}
\end{eqnarray}
for $g'<1$ since $f < r_0$, that is, in the weak field limit.

\begin{figure}[tbp]
    \includegraphics[width=0.45\textwidth]{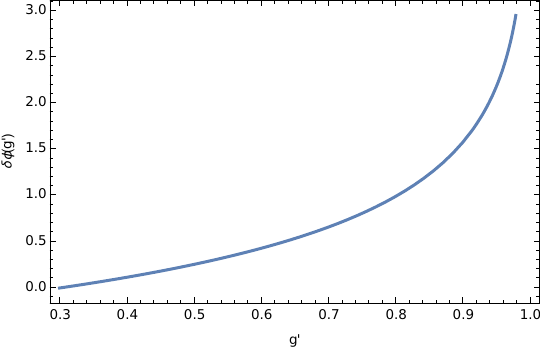}
    \hfill
    \includegraphics[width=0.45\textwidth]{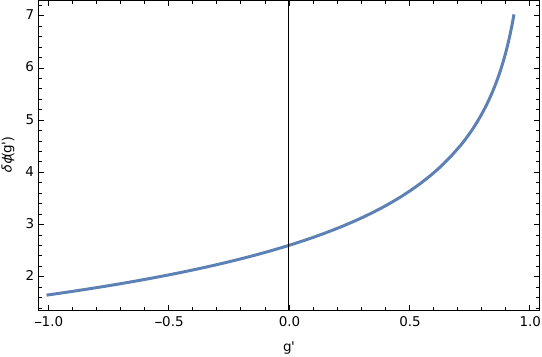}
    \caption{\label{fig: 8} Deflection angle for $\alpha_0=1$ (left one) and $\alpha_0=1/2$ (right one). Here, $a=2.2$, $b=2.3$, $\beta'=1.5$.}
\end{figure}

Therefore, the deflection angle from (\ref{f9}) using (\ref{f12}) becomes
\begin{eqnarray}
    \delta\phi (g')=\frac{2\,\beta'}{\alpha_0\,\sqrt{b^2-a^2+\beta'^2}}\,K(g')-\pi.
    \label{f13}
\end{eqnarray}
We plot the deflection angle given by equation (\ref{f13}) as a function of $g'$ in Figure 8, considering both cases: without the effects of the global monopole parameter $\alpha_0=1$, and with the effects of the global monopole parameter $\alpha_0 \neq 1$. It is observed that the deflection angle is significantly increased in the presence of the global monopole, indicating the substantial influence of the global monopole on the deflection of light.

As the parameter $g'$ always falls within the range $0 < g' < 1$, the deflection angle given by equation (\ref{f13}) can be expressed as follows:
\begin{eqnarray}
    \delta\phi=\frac{\pi\,\beta'}{\alpha_0\,\sqrt{b^2-a^2+\beta'^2}}\Bigg[1+\frac{(b^2-a^2)/4}{\Big(b^2-a^2+\beta'^2\Big)}+\frac{9(b^2-a^2)^2/64}{\Big(b^2-a^2+\beta'^2\Big)^2}+\mathcal{O} (g')^6\Bigg]-\pi.
    \label{f14}
\end{eqnarray}

It is evident that the deflection angle for null geodesics is influenced by global monopole characterized by the parameter $\alpha_0$ and undergoes modifications. When $b^2=a^2$, space-time (\ref{aa6}) corresponds to Klinkhamer defect wormhole with a global monopole. In that case, the expression for the deflection angle, derived from equation (\ref{f14}), can be written as $\delta\phi=\pi\,\left(\frac{1}{\alpha_0}-1\right)>0$.

\subsection{Gravitational lensing by wormhole with a global monopole for $M \neq 0$}

In this part, we study the deflection of photon ray for nonzero mass, $M \neq 0$ in the space-time geometry (\ref{aa1}). By considering Eqs. (\ref{f3}) and (\ref{f4}) for the null geodesics for the case $M \neq 0$, we obtain the following expression: 
\begin{eqnarray}
    \frac{d\phi}{dr}=\frac{\dot{\phi}}{\dot{r}}=\frac{\beta'}{\alpha_0}\,\frac{1}{\sqrt{(r^2-f^2)^2-\beta'^2\,\Big(1-\frac{2\,M}{r}\Big)\,(r^2-f^2)}}.
    \label{g1}
\end{eqnarray}
Therefore, the deflection angle is given by \cite{aa28}
\begin{equation}
    \delta\phi=\Delta\phi-\pi,
    \label{g2}
\end{equation}
where
\begin{equation}
    \Delta\phi=\frac{2}{\alpha_0}\,\int^{\infty}_{\tilde{r}}\,\frac{dr}{\sqrt{r^2-f^2}\,\sqrt{\Big(\frac{r^2-f^2}{\beta'^2}\Big)-\Big(1-\frac{2\,M}{r}\Big)}}.
    \label{g3}
\end{equation}
Here $\tilde{r}$ is obtained by setting $\frac{dr}{d\tau}|_{r=\tilde{r}}=0$ for null geodesics, and one will find that the impact parameter in terms of this point will be $\beta' (\tilde{r})=\sqrt{\tilde{r}^2-f^2}\,\Big(1-\frac{2\,M}{\tilde{r}}\Big)^{-1/2}$. From equation (\ref{g3}), we obtain 
\begin{equation}
    \Delta\phi=\frac{2}{\alpha_0}\,\int^{\infty}_{\tilde{r}}\,\frac{dr}{\sqrt{r^2-f^2}\,\sqrt{\Big(\frac{r^2-f^2}{\tilde{r}^2-f^2}\Big)-\Big(1-\frac{2\,M}{r}\Big)}}.
    \label{g4}
\end{equation}

If we consider $b^2=a^2$, as discussed in {\bf case A} in {\it section 4}, space-time (\ref{aa1}) becomes a non-vacuum solution of the field equations. Therefore, for $b^2=a^2$, we have $f \to 0$, and setting the impact parameter $\beta' (\tilde{r}_0)=\tilde{r}_0\,\Big(1-\frac{2\,M}{\tilde{r}_0}\Big)^{-1/2}$, where $\tilde{r}_0>2\,M$. In that case, we obtain from (\ref{g4}) the following expression
\begin{eqnarray}
    \Delta\phi&=&\frac{2}{\alpha_0}\,\int^{\infty}_{\tilde{r}_0}\,\frac{dr}{r\,\sqrt{\Big(\frac{r}{\tilde{r}_0}\Big)^2\Big(1-\frac{2\,M}{\tilde{r}_0}\Big)-\Big(1-\frac{2\,M}{r}\Big)}}=\frac{1}{\alpha_0}\,\Delta\phi_{Sch}
    \label{g5}
\end{eqnarray}
which is equals to $\frac{1}{\alpha_0}$ times the value obtained in the background of Schwarzschild-type black hole solution \cite{SW}. We see that the presence of a global monopole increases the deflection angle for null geodesics compared to the known expression in the literature.

\section{Conclusions}

This study investigates the impact of topological defects produced by cosmic strings and global monopoles on the effective potential of massless and/or massive particles in the context of Schwarzschild-Klinkhamer-type (SK) wormhole, a non-vacuum solution of the field equations that satisfies the weak and the null energy conditions. By deriving the effective potential in the presence of cosmic strings and global monopoles, we demonstrated that the topological defects modified the effective potential of the system, and induced shifts in the results. Specifically, the effective potential increases under the influence of cosmic strings and decreases due to the presence of global monopoles. We visually illustrate the effects of cosmic string on the effective potential for null and time-like geodesics through graphical representations (figures 1-4). Also, we visually illustrate the effects of global monopole on the effective potential for null geodesic by figures 6-7. These plots provide a clear understanding of how the presence of the topological defects alters the behavior of particles moving along geodesics in this SK-wormhole background.

Furthermore, we determine the deflection angle for null geodesics in this wormhole background with a cosmic string and global monopole, focusing on the special case where $M=0$ and later on $M\neq 0$. Our analysis reveals that the deflection angle is influenced by the cosmic string and global monopole, resulting in an increased deflection compared to the case without the topological defects. We have presented graphical representations of this deflection angle for the condition $b^2 > a^2$, and $a>0$ (figures 5 \& 8) and observed the significant impact of cosmic strings and global monopoles on this angle of deflection for null geodesics.

In summary, this comprehensive analysis investigated the behavior of geodesics in Schwarzschild-Klinkhamer-type wormhole, taking into account the presence of topological defects such as cosmic strings and global monopoles. By deriving and analyzing the effective potential, we gain valuable insights into the modifications and effects induced by these topological defects within the wormhole geometry. Additionally, we determined the deflection angle for photon rays in both scenarios, highlighting the impact of cosmic strings and global monopoles on this angle. Overall, this study provides a deeper understanding of how topological defects can modify the behavior of geodesics and the deflection of light within the considered wormhole geometries.

\end{document}